\documentclass{PoS}

\title{$\gamma\gamma \to \gamma \gamma$ scattering in ultrarelativistic UPC}

\ShortTitle{$\gamma\gamma \to \gamma \gamma$ scattering in ultrarelativistic UPC}

\author{\speaker{Antoni Szczurek\thanks{A footnote may follow.}}\\
        Institute of Nuclear Physics Polish Academy of Sciences, PL-31342 Krakow, Poland\footnote{Also at \textit{Faculty of Mathematics and Natural Sciences, University of Rzeszow, ul. Pigonia 1, 35-310 Rzeszow}}\\
        E-mail: \email{Antoni.Szczurek@ifj.edu.pl}}

\author{Mariola K{\l}usek-Gawenda\\
        Institute of Nuclear Physics Polish Academy of Sciences, PL-31342 Krakow, Poland\\
        E-mail: \email{Mariola.Klusek@ifj.edu.pl}}

\abstract{We discuss diphoton semi(exclusive) production in ultraperipheral 
lead-lead collisions at energy of $\sqrt{s_{NN}}$ = 5.5 TeV (LHC)
and in proton-proton collisions at $\sqrt{s_{pp}}$ = 7 TeV (LHC) 
and $\sqrt{s_{pp}}$ = 100 TeV (FCC).
The nuclear calculations are based on equivalent photon approximation 
in the impact parameter space. 
The cross sections for elementary $\gamma\gamma \to \gamma\gamma$ subprocess 
are calculated including three different mechanisms: 
box diagrams with leptons and quarks in the loops, 
a VDM-Regge contribution with virtual intermediate hadronic excitations 
of the photons and the two-gluon exchange contribution. 
We get relatively high cross sections in heavy ion collisions. 
This opens a possibility to study the light-by-light 
(quasi)elastic scattering at the LHC. 
We find that the cross section for elastic $\gamma\gamma$ scattering
could be measured in the lead-lead collisions for the diphoton invariant
mass up to $W_{\gamma\gamma} \approx$ 15 - 20 GeV. 
Our Standard Model predictions are compared to a recent ATLAS
experimental result.
We present differential distributions for PbPb$\to$PbPb$\gamma\gamma$ 
and pp$\to$pp$\gamma\gamma$ reaction. 
}

\FullConference{EPS-HEP 2017, European Physical Society conference on High Energy Physics\\
		5-12 July 2017\\
		Venice, Italy}

\begin{document}

\section{Introduction}

In classical Maxwell theory photons/waves/wave packets do not interact. 
In contrast, in quantal theory they can interact via quantal fluctuations. 
So far only inelastic processes, i.e. production of hadrons or jets 
via photon-photon fusion could be measured. 
It was realized only recently that ultraperipheral collisions (UPC) of 
heavy-ions can be also a good place where photon-photon elastic
scattering could be tested experimentally \cite{ES_LbL,KGLSz2016}.
Our calculation \cite{KGLSz2016} gave an order of magnitude
larger cross section than in the earlier calculation \cite{ES_LbL}
(see also erratum \cite{ES_LbL_erratum} made by the authors of \cite{ES_LbL}).
The theoretical predictions of both groups became a motivation 
for ATLAS experimental group to perform corresponding experimental studies.
Recently the ATLAS Collaboration observed 13 events for light-by-light 
scattering in Pb-Pb UPC \cite{ATLAS}. 
We shall discuss our results and future possibilities.

\section{$\gamma\gamma \to \gamma\gamma$ elementary cross section}

\begin{figure}[!h]
\centering
\includegraphics[scale=0.25]{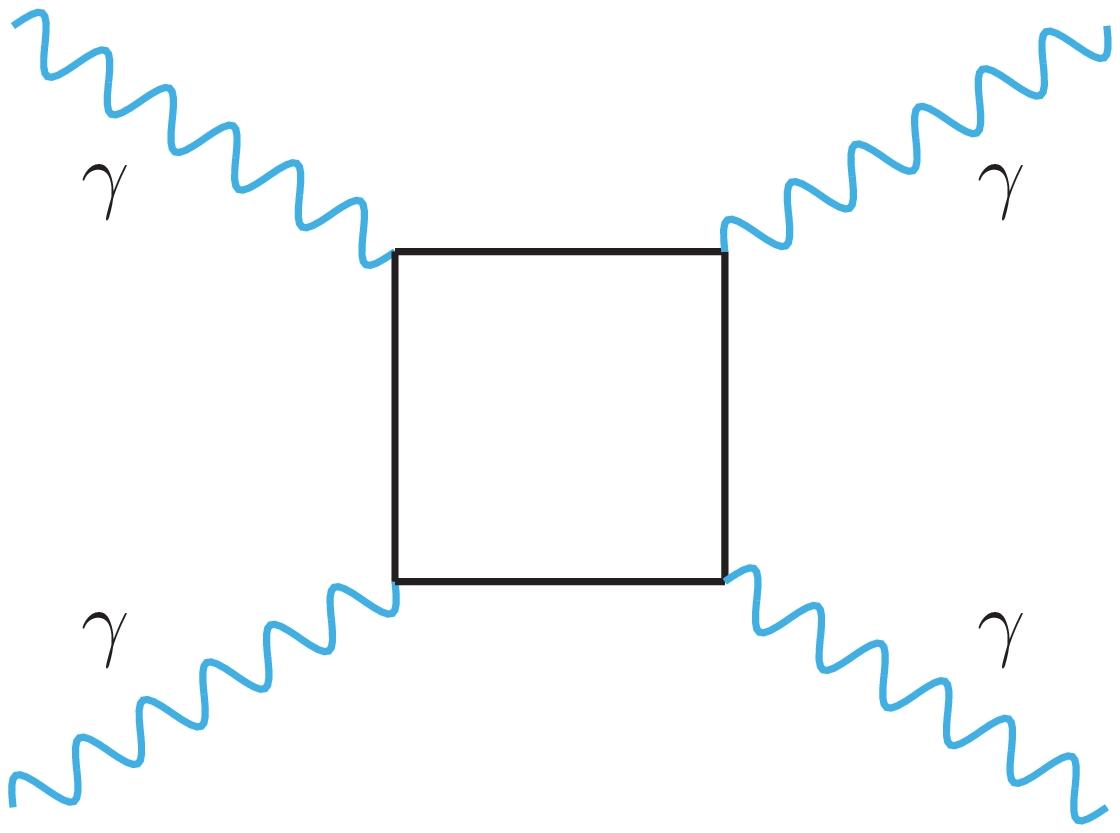}
\includegraphics[scale=0.25]{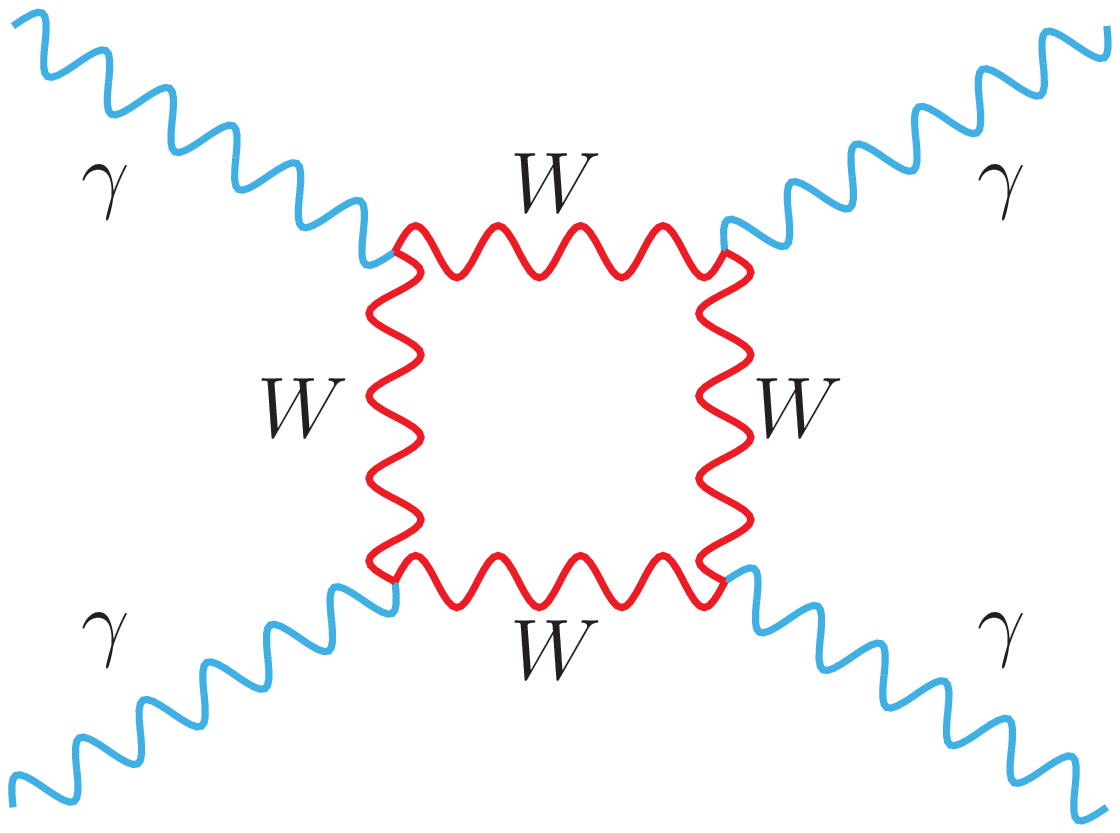}
\includegraphics[scale=0.35]{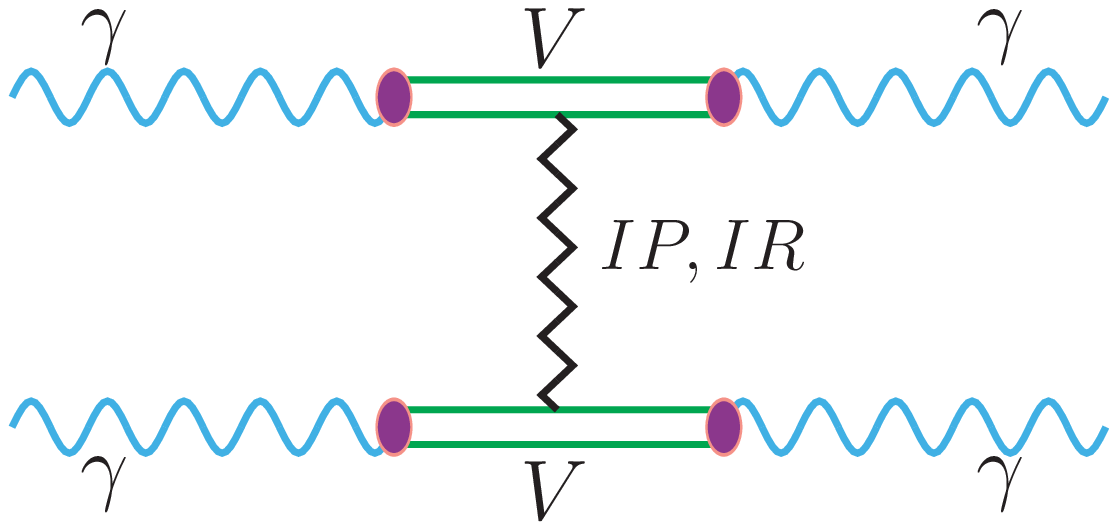}
\includegraphics[scale=0.35]{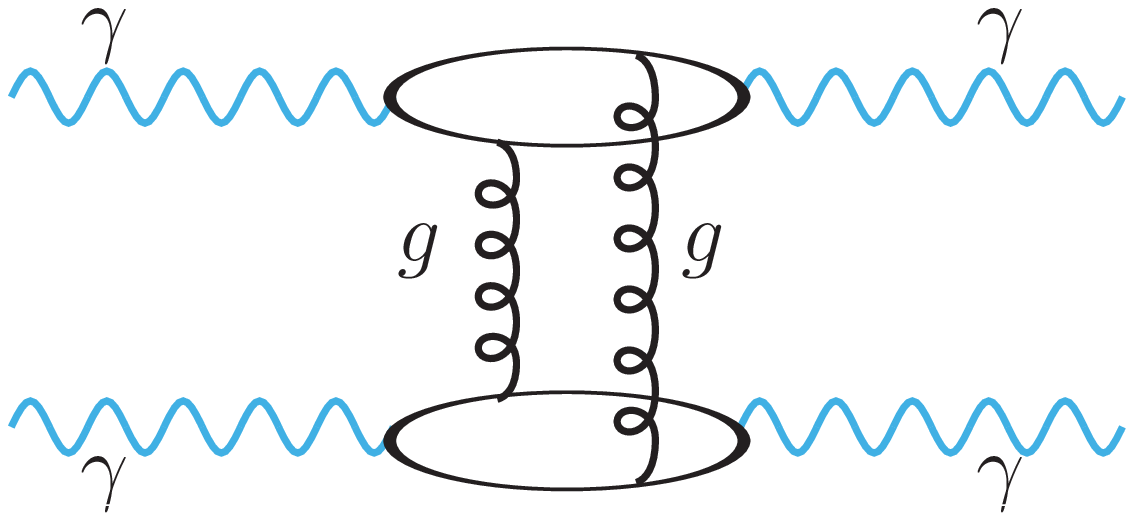}
\caption{Light-by-light scattering mechanisms with 
the lepton and quark loops (first panel) and for the intermediate 
$W$-boson loop (second panel). 
The third panel represents VDM-Regge mechanism and the last panel is for
two-gluon exchange.}
\label{fig:diagrams_elementary}
\end{figure}

The lowest order QED mechanisms with elementary particles are shown
in two first diagrams of Fig.~\ref{fig:diagrams_elementary}. 
The first diagram is for lepton and quark loops and it dominates
at lower photon-photon energies ($W_{\gamma\gamma}<2m_W$) 
while the next diagram is for the $W$ (spin-1) boson loops and it becomes 
dominant at higher photon-photon energies (\cite{Bardin2009,Lebiedowicz2013}). 
The one-loop box amplitudes were calculated by using
the Mathematica package {\tt{FormCalc}} and the {\tt{LoopTools}} library.
We obtained good agreement when confronting our result with those in 
\cite{Bardin2009,Jikia1993,Bern2001}.
In Ref.~\cite{Bern2001} the authors considered
the QCD and QED corrections 
to the one-loop fermionic contributions in the 
limit ($\hat{s},|\hat{t}|,|\hat{u}| \gg m_f^2$). 
The corrections are quite small numerically 
so the LO computations are satisfactory.
In the last two diagrams of Fig.~\ref{fig:diagrams_elementary} we 
show processes that are the same order in $\alpha_{em}$ but higher order
in $\alpha_s$. They were discussed only by our group.
The third diagram presents situation where
both photons fluctuate into virtual vector mesons ($\rho, \omega, \phi$). 
The last diagram shows two-gluon exchange mechanism which is formally 
three-loop type.
Its contribution to the elastic scattering of photons at high energies 
was considered first in the pioneering work \cite{Gin87}. 
Indeed in the limit where the Mandelstam variables of 
the $\gamma \gamma \to \gamma \gamma$ process satisfy
$\hat s \gg -\hat t$, $-\hat u$, 
major simplifications occur and this process becomes tractable. 
In our treatment, we include finite fermion masses, 
the full momentum structure in the loops as well as all helicity 
amplitudes \cite{KGSSz2016}.

\section{Light-by-light scattering in UPC}

A nuclear cross section is calculated in 
equivalent photon approximation (EPA) in the impact parameter space. 
The total (phase space integrated) cross section is expressed through 
the five-fold integral (for more details see e.g.~\cite{KG2010})
compared to two-dim integration in \cite{ES_LbL} 
\begin{equation}
\sigma_{A_1 A_2 \to A_1 A_2 \gamma \gamma}\left(\sqrt{s_{A_1A_2}} \right) =
\int \sigma_{\gamma \gamma \to \gamma \gamma} 
\left(W_{\gamma\gamma} \right)
N\left(\omega_1, {\bf b_1} \right)
N\left(\omega_2, {\bf b_2} \right)  \, S_{abs}^2\left({\bf b}\right)
\mathrm{d}^2 b \, \mathrm{d}\overline{b}_x \, \mathrm{d}\overline{b}_y \, 
\frac{W_{\gamma\gamma}}{2}
\mathrm{d} W_{\gamma\gamma} \, \mathrm{d} Y_{\gamma \gamma} \;,
\label{eq:EPA_sigma_final_5int}
\end{equation}
where $W_{\gamma\gamma}$ 
and $Y_{\gamma \gamma}$ 
is invariant mass and rapidity of the outgoing $\gamma \gamma$ system. 
Energy of photons is expressed through 
$\omega_{1/2} = W_{\gamma\gamma}/2 \exp(\pm Y_{\gamma\gamma})$.
$\bf b_1$ and $\bf b_2$ are impact parameters 
of the photon-photon collision point with respect to parent
nuclei 1 and 2, respectively, 
and ${\bf b} = {\bf b_1} - {\bf b_2}$ is the standard impact parameter 
for the $A_1 A_2$ collision.
In practical calculation we use 
$S_{abs}^2 = \theta(|{\bf b}_1 - {\bf b}_2| - 2 R_A)$.
The photon flux ($N(\omega,b)$) is expressed through a nuclear charge
form factor of the nucleus. In our calculations we use a realistic 
form factor which is a Fourier transform of the charge distribution 
in the nucleus.
More details can be found e.g. in \cite{KGLSz2016,KG2010}.

\begin{figure}
\centering
\includegraphics[scale=0.25]{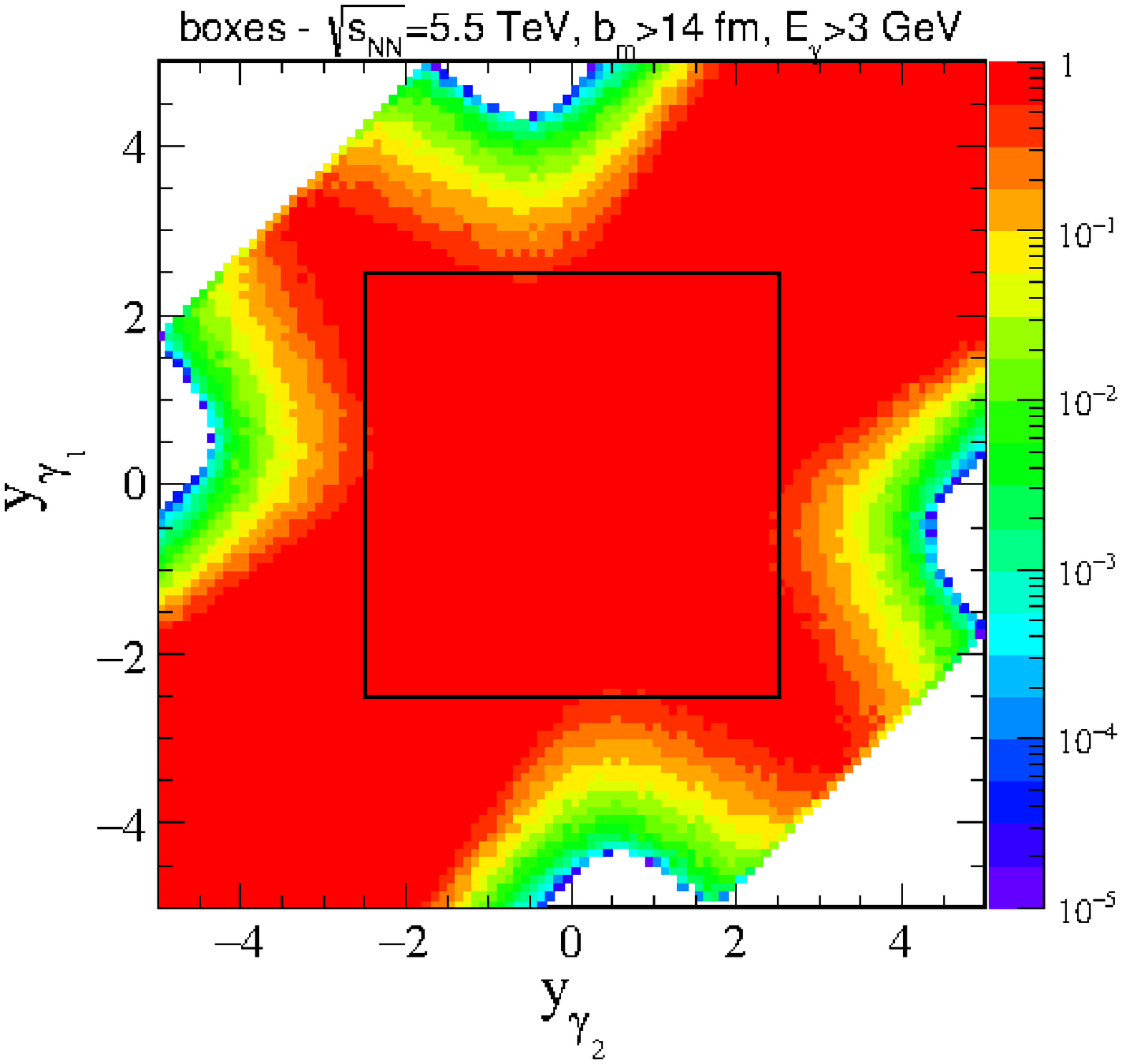}
\includegraphics[scale=0.25]{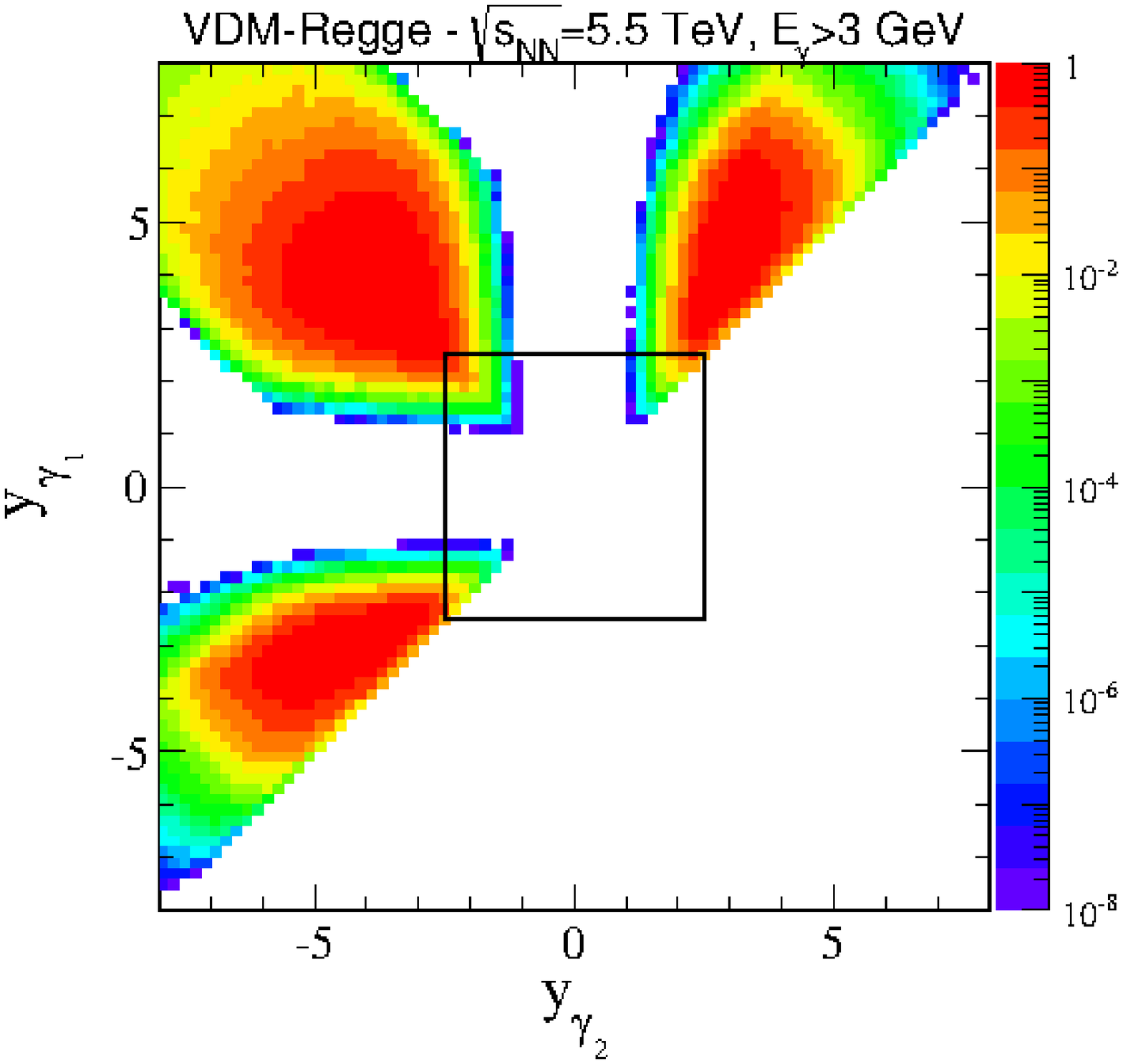}
\includegraphics[scale=0.24]{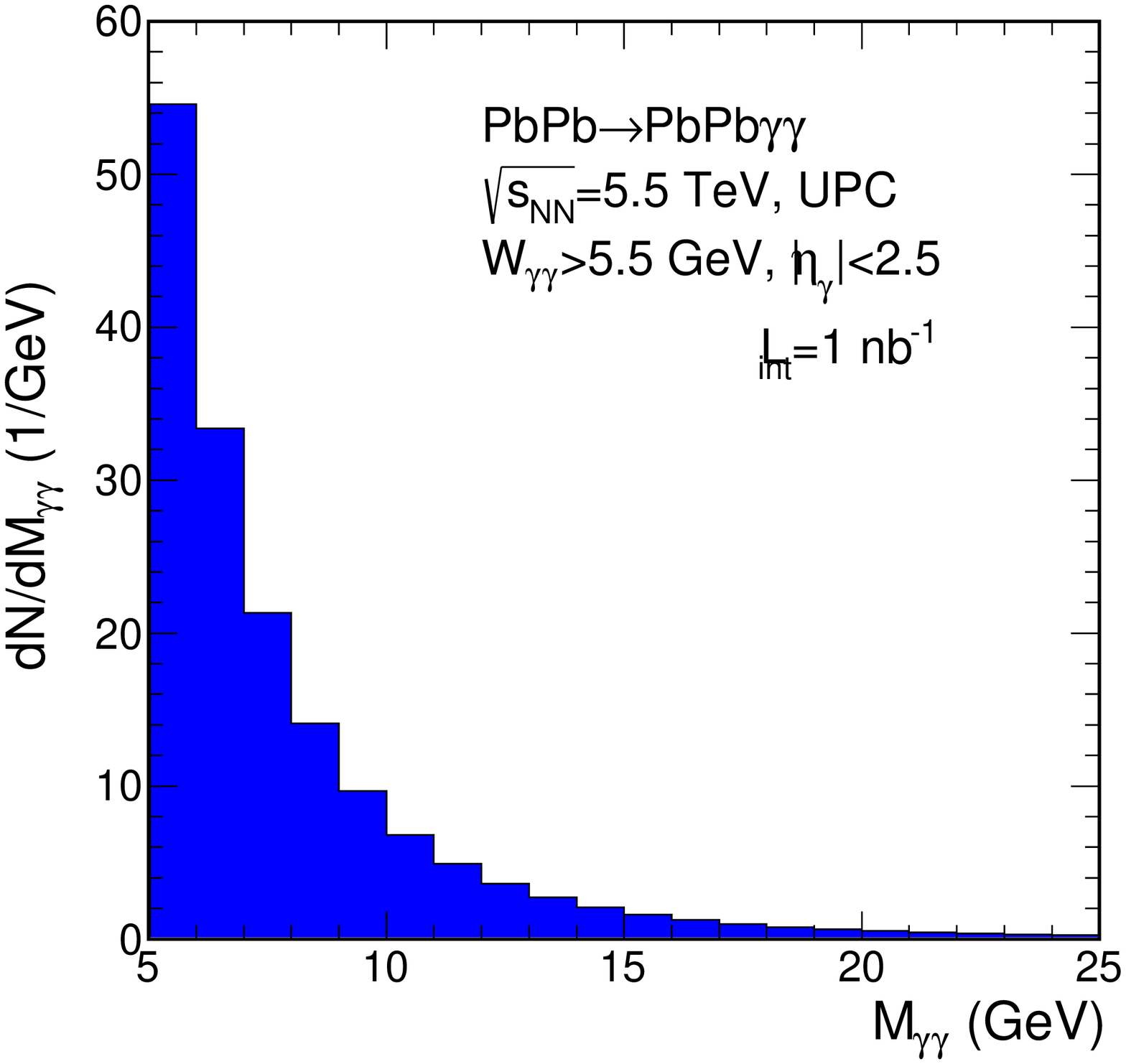}
\caption{\label{fig:dsig_dy1dy2}
Contour representation (first two panels) of two-dimensional 
($\mathrm{d}^2 \sigma / \mathrm{d} y_{\gamma_1} \mathrm{d} y_{\gamma_2}$ in nb) 
distribution in rapidities 
of the two photons in the laboratory frame for box (left panel) 
and VDM-Regge (middle panel) contributions. Nuclear calculations were
done for $\sqrt{s_{NN}}= 5.5$ TeV. The last panel shows distribution 
of expected number of counts in $1$~GeV bins 
for cuts on $W_{\gamma\gamma}>5.5$ GeV and $\eta_\gamma<2.5$.}
\end{figure}

In our recent papers we tried to answer the question whether the
reaction can be measured with the help of LHC detectors.
Then we have to generalize Eq.~(\ref{eq:EPA_sigma_final_5int}) 
by adding extra integration over additional parameter related 
to angular distribution for the subprocess \cite{KGLSz2016}.
The two first panels of Fig.~\ref{fig:dsig_dy1dy2} show 
two-dimensional distributions in photon rapidities. 
The calculation was done at the LHC energy
$\sqrt{s_{NN}}=5.5$ TeV. Here we imposed cuts on energies of photons 
in the laboratory frame ($E_{\gamma}>3$~GeV). 
We obtain very different distributions for box and VDM-Regge mechanisms. 
In both cases the influence of the imposed cuts is significant.
In the case of the VDM-Regge
contribution we observe non continues behaviour 
which is caused by the strong transverse momentum dependence of 
the elementary cross section (see Fig.~4 in Ref.~\cite{KGLSz2016})
which causes that some regions in the two-dimensional space are almost 
not populated. 
The second half of the ($y_{\gamma_1},y_{\gamma_2}$) space 
can be obtained from the symmetry around the $y_{\gamma_1}=y_{\gamma_2}$
diagonal.
Clearly the VDM-Regge contribution does not fit
to the main detector ($-2.5<y_{\gamma_1},y_{\gamma_2}<2.5$) and extends 
towards large rapidities.
In this case we show much broader range of rapidity
than for the box component. We discover that maxima
of the cross section associated with the VDM-Regge mechanism are at
$|y_{\gamma_1}|,|y_{\gamma_2}| \approx$ 5. Unfortunately this is below 
the limitations of the ZDCs for ATLAS ($|\eta| > 8.3$) 
and CMS ($|\eta| > 8.5$).\\
The last panel of Fig. \ref{fig:dsig_dy1dy2} presents
the numbers of counts in the $1$ GeV intervals expected for assumed
integrated luminosity: $L_{int}=1$~nb$^{-1}$ typical for UPC at the LHC.
In this calculation we imposed cuts on photon-photon energy and
(pseudo)rapidities of both photons.
With the assumed luminosity one can measure invariant mass distribution up to 
$M_{\gamma \gamma} \approx 15$ GeV. 

We studied the mechanism of elastic photon-photon scattering also
in $pp \to pp \gamma\gamma$ reaction. 
In our calculations we neglect the gap survival factor.
The cross section of $\gamma\gamma$ production in proton-proton collisions 
takes the simple parton model form
\begin{equation}
\frac{\mathrm{d}\sigma}{\mathrm{d}y_1 \mathrm{d}y_2 \mathrm{d}^2p_t} = \frac{1}{16\pi^2 \hat{s}^2}
x_1 \gamma^{(el)} (x_1)
x_2 \gamma^{(el)} (x_2) 
\overline{\left|{M}_{\gamma\gamma \to \gamma \gamma}\right|^2} \;.
\label{eq:pp}
\end{equation}
Here $y_{1/2}$ is the rapidity of final state photon, 
$p_t$ is the photon transverse momentum
and $x_{1/2} = p_t/\sqrt{s} \left( \exp\left( \pm y_1 \right) + \exp\left( \pm y_2 \right) \right)$.
The calculation for proton-proton collisions are done as usually in the
parton model with elastic photon distributions expressed 
in terms of proton electromagnetic form factors.  
Detailed description of the cross section for pp$\to$pp$\gamma\gamma$ reaction
can be found in our paper \cite{KGSSz2016}.

\begin{figure}
\centering
\includegraphics[scale=0.25]{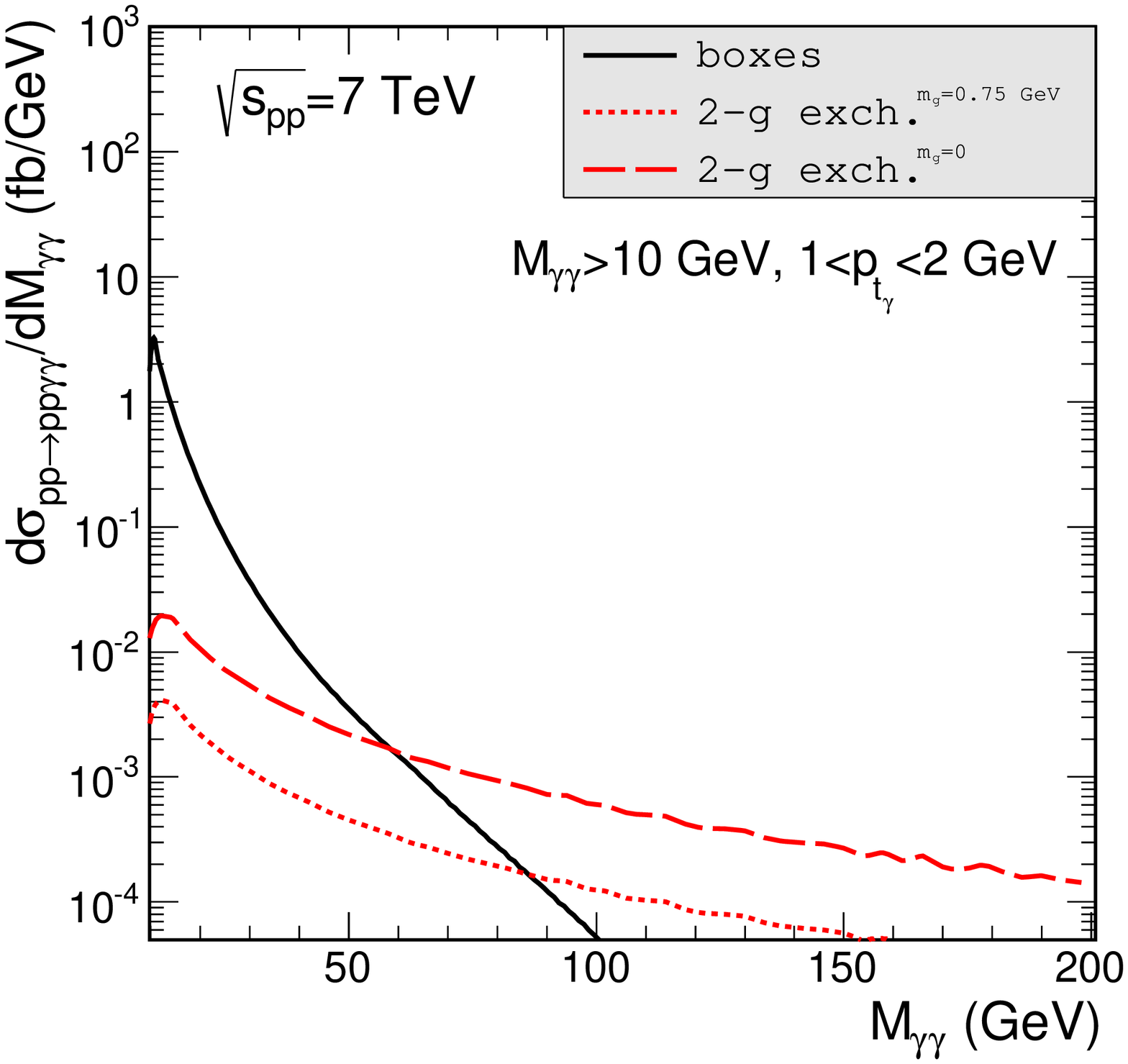}
\includegraphics[scale=0.25]{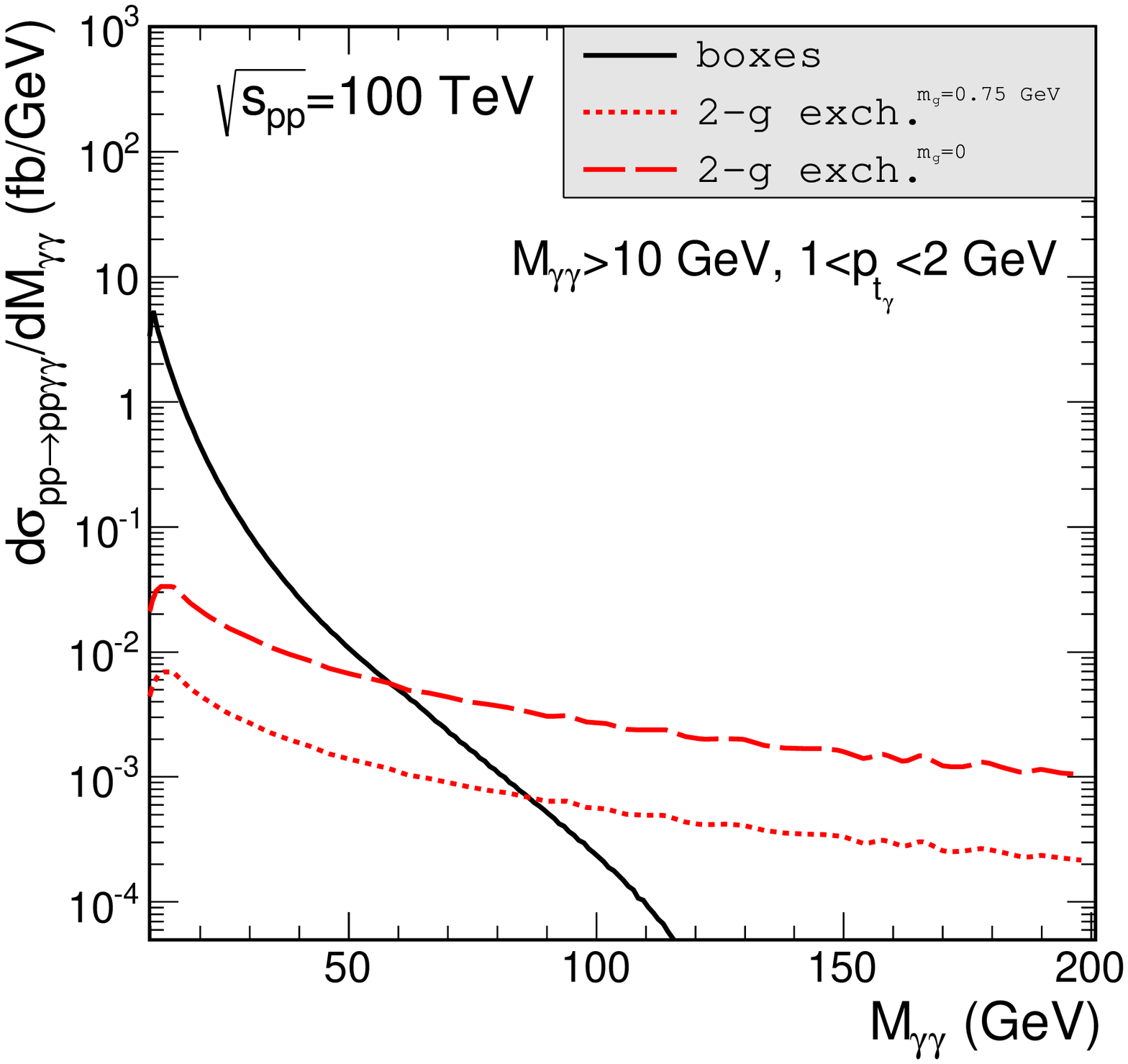}
\caption{\label{fig:dsig_dW_LHC_FCC}
Distribution in invariant mass of the produced photons for $\sqrt{s_{pp}} = 7$~TeV (LHC) 
and $\sqrt{s_{pp}} = 100$ TeV (FCC) for cuts on photon transverse momenta specified in the figure legend. No cuts on photon rapidities are applied here.}
\end{figure}

The nuclear distribution in the diphoton invariant mass is shown 
in Fig.~\ref{fig:dsig_dW_LHC_FCC}. 
The two-gluon distribution starts to dominante over the box contribution
only above $M_{\gamma\gamma} > 50$ GeV for $1$ GeV $< p_t < 2$ GeV. 
However, the cross section in this region is rather small.
The situation for the LHC (left panel) and for Future Circular 
Collider (FCC) energy (right panel) is rather similar. 
The dominance of the two-gluon exchange over the box contribution takes
place more or less at the same diphoton invariant masses.

\section{Conclusion}

In our recent papers we studied how to measure elastic photon-photon 
scattering in ultrarelativistic lead-lead UPC as well 
as proton-proton collisions. 
The cross section for photon-photon scattering was calculated taking 
into account well known box diagrams with elementary standard 
model particles (leptons and quarks), 
a VDM-Regge component as well as a two-gluon exchange,
including massive quarks, all helicity configurations of photons and massive and massless gluon. 
For the PbPb$\to$PbPb$\gamma\gamma$ reaction 
we identified regions of the phase space where the two-gluon contribution 
should be enhanced relatively to the box contribution. 
The region of large rapidity difference between the two emitted photons 
and intermediate transverse momenta 1 GeV < $p_t$ < 2-5 GeV seems
optimal in this respect.

This year the ATLAS Collaboration published a new result \cite{ATLAS} 
for light-by-light scattering in quasi-real photon interactions 
which come from ultraperipheral lead-lead collisions at 
$\sqrt{s_{NN}}$= 5.02~TeV. 
The measured fiducial cross section which includes limitation 
on photon transverse momentum, photon pseudorapidity, diphoton invariant mass, 
diphoton transverse momentum and diphoton accoplanarity, 
were measured to be 70 $\pm$ 20 (stat.) $\pm$ 17 (syst.) nb. 
This result is compatible with the value of 49 $\pm$ 10 nb predicted by us for
the ATLAS cuts and experimental luminosity.


\end{document}